\documentclass[aps,prd,showpacs,amssymb,nofootinbib,twocolumn]{revtex4}
\usepackage{graphicx}
\usepackage{amsmath}
\usepackage{xcolor}

\def\half {{1\over 2}}

\def\bea{\begin{eqnarray}}
\def\eea{\end{eqnarray}}
\def\ket#1{\vert #1 \rangle}
\def\bra#1{\langle #1 \vert}

\def\change#1{{\color{blue}{#1}}}
\def\change#1{{\color{black}{#1}}}
\def\erase#1{{\color{red}{#1}}}
\def\erase#1{{}} 

\def\sqr#1#2{{\vcenter{\vbox{\hrule height.#2pt
   \hbox{\vrule width.#2pt height#1pt \kern#1pt
      \vrule width.#2pt}
   \hrule height.#2pt}}}}

\begin{document}
\title{ Black Hole --Entropy Container or Creator \\
 \change{using a simple Black Hole model and a simple linear
 amplifier model}}

\author{W. G. Unruh}
\affiliation{
Dept.~of Physics and Astronomy,
University of B.~C.\\
Vancouver, Canada V6T 1Z1\\
Hagler Fellow and IQSE Distinguished Researcher\\
IQSE, TAMU-4242
Texas A\&M Univ.\\
College Stn, Tx, USA 77843-4242\\
email: unruh@physics.ubc.ca}

~

~

\begin{abstract}

Do Black Holes possess entropy or do they create it? \erase{The dominant assumption i
that they possess entropy, and a they evaporate that entropy is emitted and
decreases.}
\change{The dominant assumption in most work on Black Hole
entropy is that Black Holes possess entropy (created via some unknown
non-linear process within the horizon of the Black Hole)
and as they evaporate, that entropy is slowly emitted and
the Black Hole's entropy slowly decreases. This is the ``lump of
coal" model.
}In this paper I 
\change{present
}a model of a
linear  amplifier, in which I
argue that the amplifier has 
\change{no
}entropy and yet it emits entropy in
the process of 
\change{its
}operation. 
\change{This entropy is
"entanglement entropy".
}This model is closely related to
behaviour of Black Holes, resulting, 
\change{in answer to the question raised in
the 
title,
}in the answer that Black Holes do not have entropy, but nevertheless 
create and emit entropy. 
\change{The  total entropy emitted during the
lifetime of the Black Hole
}is  the same
as the usual expression proportional to 
\change{1/4 the area of the horizon of the
Black Hole.
} 

\end{abstract}

\maketitle
\section{Hawking evaporation}
Over 50 years ago, 
\change{Hawking\cite{Hawking} 
}calculated that Black Holes were not black, but
rather emitted radiation with a thermal spectrum whose temperature was
proportional to the inverse mass of the Black Hole 
\change{
for a Schwarzschild black
hole.
\bea
T={\hbar c^3 \over {8{\pi}K_BGM}}
\eea
}
By using the thermodynamic
relation:
\bea
TdS=dE,
\eea
\erase{where T is the temperature of the emitted radiation, E is the mass of the
Black Hole times $c^2$, and $dS$ is the entropy emitted}
\change{ he ascribed
an entropy of 
\bea
S={ {4K_B\pi G M^2\over {c\hbar}}}={K_B c^3\over 4\hbar G}A=K_B {1\over
4} { \bf A},
\eea
where $A$ is the surface area in regualr units, and$\bf A$ is the surface area of the
horizon of the Black Hole in Planck units. Note that this is the
entropy emitted by the Black Hole integrated over the lifetime of the
Black Hole, and $T$ is the temperature ( which depends on the energy of
the Black Hole). Since the area of the horizon is an aspect of
the Black Hole, it strongly suggested to most people that the total entropy is an
aspect of the Black Hole.
}
\erase{In his calculation, Hawking
found an emitted spectrum of any quantum field with a temperature equal to
temperature
\bea
T={\hbar c^3 \over {8{\pi}K_BGM}}
\eea
where $K_B$ is Boltzmann's constant
This gives an entropy of 
\bea
S={ {4K_B\pi G M^2\over {c\hbar}}}={K_B c^3\over 4\hbar G}A
\eea
where $A$ is the area of the horizon.}
\change{ However, the total entropy is defined as the total entropy radiated 
to infinity by the
Black Hole in its lifetime, not the entropy contained in the Black Hole
after its formation.
}

What is this entropy? The usual argument is that this is an entropy possessed
by the Black Hole, and as the Black Hole evaporated, this entropy is depleted
by the radiation emitted by the Black Hole.  The model is either the
Planck black-box radiation, where the entropy is essentially that of the
oscillators in the inside walls of the oscillator, which,
\change{because of
} the dipole
radiation produced by  these  oscillators, produces the entropy
\erase{in the} 
\change{carried away by the
}
Electromagnetic radiation. Or, equivalently, it can be modelled as the
entropy, like that of 
of a lump of coal,
heated up by some incoming pure-state radiation, which, by non-linear
interactions of the constituents of the coal, creates entropy 
\change{within the coal.
}\erase{It}\change{The hot coal} then slowly emits that entropy
(and energy). But this analogy leads to the question of where that entropy is stored in
the Black Hole. Some ideas \erase{is}\change{are} that the entropy is related
\change{to degrees of freedom
} of the surface of
the Black Hole represented by the horizon, and that each square Planck
area somehow stores 
\change{1/4 of a
}unit of entropy.
\change{Alternatively,
}the entropy is contained inside the
horizon of the Black Hole and tunnels
\cite{tunnel} through the horizon
\change{(despite the
the fact that the 
}inside and outside are causally disconnected regions).
\change{Alternatively,
}quantum gravity
causes the horizon to shake producing the radiation. Or, 
in string theory\cite{string}, the horizon is represented by some sort of null d-brane
with the strings penetrating the horizon represent the degrees of freedom
which are excited as vacuum fluctuation and represent the entropy of the black
holes. All have problems. Since the right answer is 
\change{supposedly
}known (the Hawking
entropy), 
\change{the question
} can be answered by one means or the other. An  example of such a
problem is that, since each
field radiates with the above thermal spectrum, the entropy emitted will be
proportional to the log of the number of quantum fields in one's theory,
which could rapidly far exceed the  Hawking entropy. \erase{The answer is to
argue}
\change{ Perhaps 
} $G$, Newton's constant, could be renormalized in quantum gravity
to \erase{ above
 } 
\change{cancel that
}logarithmic divergence. However, all 
\change{
suggestions
}have the problem that one seems to be
getting more out of the theory than one should, 
\change{or that one requires new
theories of physics to explain the entropy
}.

\section{Analogies}

In my career I have come across a number of similar situations. One was the
Bohr-Einstein weighing-of-energy debate where Bohr's answer seemed to demand
that General Relativity must be true if quantum mechanics is to be consistent.
G. Opat and I\cite{WeighingofEnergy} showed that one  only needed Einstein's
assumption (namely that energy has weight) to show the consistency of
quantum mechanics.\erase{ While General Relativity is based on the  assumption that
energy has weight, any
theory which contained that result  would also lead to the same consistency
that quantum mechanic's obedience of the time-energy uncertainty could be
tested in the way Einstein suggested.}

Geroch  argued \erase{against} that Bekenstein's identification of entropy with the area of
a Black Hole must be wrong. 
\change{He presented a model
} \erase{ was disproven by a} heat engine which extracted  energy by
lowering a box of entropy toward the black
hole\erase{--  could be used violate the Second Law.} 
\change{while
extracting the content's energy via the tension in the rope.
}
Bekenstein \erase{then}
\change{answered
} by  postulating a new law of physics, that there exists \erase{ a new law
of physics that there exists} an  upper bound of the entropy to
energy ratio of the contents of box, which would save the Second law.
Wald and I\cite{unruhwald} showed that\erase{  if one took into account the acceleration
temperature outside the accelerated box,} Archimedes' buoyancy principle
\change{ and the acceleration temperture would save the Second Law.
}\erase{ reduce the
work extracted as one lowered the box, so that no new laws
of physics were needed  to show this heat engine to obeys the Second law. }

\change{Those and other examples convinced me that
}should not in general need to introduce new laws of physics to save
apparent paradoxes in physics.

Hawking's assumption, in his calculation, was that gravity was defined by a
smooth spacetime, and certainly not a quantum spacetime. It also said nothing
about the statistical nature of the entropy. The argument for the entropy
arose solely from  thermodynamic arguments.\erase{ not from any additional assumptions
about the microscopic nature of gravity. He showed that} The emission of a
thermal stream was a direct consequence of linear quantum field theory in a
Black Hole spacetime. \erase{Therefor any explanation of nature of the} The emitted entropy
should not require any additional assumptions about the nature of quantum
gravity or about the Black Hole containing  entropy.

\section{Amplifier model}

A number of years ago I presented a model of a 
\change{linear
}amplifier\cite{bhamp}. (Some of
these ideas were also considered earlier\cite{schwablthirring}). Let me
summarize 
\change{my simple model
here since it is interesting in its own right}. Consider two
quantum fields, 
$\phi$ and $\psi$, and 
\change{a single extra degree of freedom I will call a
}
harmonic oscillator with dynamic variables $p,~ q$. 
\change{The model has
linear equations in all degrees of freedom. To simplify the model,
} let me operate in 1+1
dimensional spacetime 
\change{and take  the oscillator  to have zero frequency
(i.e., be a free particle)
}, with\erase{action}
\change{Lagrangian
\bea
{\mathfrak L}&=&\half\bigg(\int_{-\infty}^0
\left[(\partial_t\phi)^2-(\partial_x\phi)^2 \right]
-\left[  (\partial_t\psi)^2-(\partial_x\psi)^2\right] \nonumber 
\\
&&+\left[(\partial_t q)^2) +2\partial_t q(\epsilon
\phi+\mu\psi)\right]\delta(x)dx\bigg) 
\eea
and}
with boundary condition on the fields that their\erase{z derivative at
$z=0+$} 
\change{x derivative at $x=0+$
} is
zero. The fields come in from \erase{$z=-\infty$}
\change{$x=-\infty$
},
interact with the oscillator at
$x=0$ and are immediately reflected back toward \erase{$z=-\infty$}
\change{$x=-\infty$.
}Furthermore,
\change{assume
}
$\mu^2<\epsilon^2$, which keeps the system stable.

Note that 
\change{in
this model,
}the $\psi$ field
has a negative Hamiltonian. This could be realised  if there is a state
of the field with a maximum value of the energy (e.g., a collection of
spin 1/2 particles with all of the particles in their maximum energy
state.) The ``vacuum" state of the $\psi$
field is an energy maximum, rather than a minimum. In a physical system, the $\psi$
system would  be non-linear for large deviations from that ``vacuum" but I
am assuming that the deviations from the ``vacuum" are small.\erase{ for the system of
interest}.

\change{This system has a conserved norm:
\bea
\langle{\sigma_1,\sigma_2}\rangle&=&{i}\left(\int [ 
{\phi_1}^*\pi_{\phi 2}-\pi_{\phi 1}^*{\phi_2}
+{\psi_1 }^*\pi_{\psi 2}-\pi_{\psi 1}^*{\psi_2}]dx \right. \nonumber\\
&&+(q^*_1 {p_2}-p_1^*{q_2})\bigg)
\eea
}
where 1 and 2 represent  any two solutions of the equations of motion, and
$\sigma$ represent the collection of
$\phi,~\pi_\phi,~\psi,~\pi_\psi,~q,~p$ for each  solution of the
equations. 
\change{In addition, $\Sigma$ is the collectiom of the quantum
Hermitian 
operators and fields $\Phi(t,x),~\Pi_\phi(t,x),~\Psi(t,x),~\Pi_\psi(t,x),~Q(t).~P(t)$ which are
Heisenberg operator solutions to the linear equations of motion.) This
norm 
is conserved for solutions, whether quantum
or classical.
}(Note that
$\pi_{\psi}=-\partial_t\psi$ because of the negativity of its
Hamiltonian/Lagrangian). Modes $\sigma$ which have
positive norms are associated with annihilation operators and the negative
norms are  
\change{(the complex conjugates of positive norm solutions)
}
associated with creation operators for the fields.
This symplectic norm is just the generalization of the Klein-Gordon norm for a single
scalar field. For the $\phi$ field the positive norms will be associated
with time dependence of $e^{-i\omega t}$ with $\omega>0$, while for
the $\psi$ field positive norm corresponds to $\omega<0$

If we take  modes such that their incoming parts (from $x=-\infty$) are of the form
\change{\bea
\phi(t,x)= \phi_{0\omega} e^{-i\omega(t-x)}\nonumber\\
\psi(t,x)=\psi_{0\omega} e^{-i\omega(t-x)},
\eea
}
then  two sets of normalized solutions for $\omega>0$ and for $x$ far to
the left, are
\bea
\phi_{0\omega}={1\over\sqrt{2\pi|\omega|}}; \psi_{0\omega}=0; ~~~[{\mathfrak
A}]\\
\psi_{0\omega}={1\over\sqrt{2\pi|\omega|}}; \phi_{\omega}=0; ~~~[{\mathfrak
B}]
\eea
The first, $\mathfrak A$, have unit positive norm (with the continuum
normalization), while the second, $\mathfrak B$,
are unit negative norm solutions. Solutions for the opposite  norms
will be the complex conjugate of these. However both ${\mathfrak A}$ and
$\mathfrak B$
have the same value of $\omega>0$.

\erase{Note that these are stationary solutions, without the oscillator degree of
freedom having independent non-zero initial values  solutions. If for example we started off with
$q,\partial_t q$ being initially non-zero and $\phi$ and $\psi$ being initially
zero,the solution would be damped   if $\mu^2<\epsilon^2$) and a run-away
solutions if $\mu^2>\epsilon^2$.}

Installing a mirror with Neumann boundary conditions just to the right of
the oscillator at 
\change{$x=\lambda\rightarrow 0$
}, the solutions for each field  are
\bea
\phi_{\omega}(t,x)&=& \lim_{\lambda=0^+}\phi_{0\omega}(e^{-i\omega
(t-x)})+e^{i\omega(t+x-2\lambda)}) \nonumber
\\
&&+\half\epsilon q_\omega
(e^{-i\omega(t+x)}+e^{-i\omega(t+x-2\lambda)}
\\
\psi_{\omega}(t,x)&=& \lim_{\lambda=0^+}\psi_{0\omega}(e^{-i\omega
(t-x)})+e^{i\omega(t+x-2\lambda)}) \nonumber\\
&&-\half\mu q_{\omega}
e^{-i\omega(t+x)}+e^{-i\omega(t+x-2\lambda)}
\eea
where $\lambda=0^+$ means were are taking the limit as $\lambda$
approaches 0 from positive values of $\lambda$. \erase{A similar equation
applies to $\psi$ }

The $q$  equation at $x=0$ is given by
\bea
\left[\omega^2 q_{\omega}- i\omega[(\epsilon(2\phi_{0\omega}
+\epsilon q_\omega+\mu(2\psi_{0\omega}-\mu q_{\omega})]e^{-i\omega t)}\right]
\eea
from which we get
\bea
q_\omega={(2i(\epsilon\phi_0+\mu\psi_0)\over(i\epsilon^2-i\mu^2+\omega)}
\eea
and with $\phi_{{\rm out}\omega}$ being the portion that goes as $e^{-\omega
x}$
\bea
\phi_{{\rm out}\omega} &=&
\\
&&-e^{-i\omega(t+x)}{(i\epsilon^2+i\mu^2-\omega)\phi_0+2i\epsilon\mu\psi_0\over(-i\epsilon^2+i\mu^2-\omega)}\\
\eea
where ``out" designates the part going as $e^{-i\omega (t+x)}$ \change{A similar
expression holds for $\psi_{{\rm out}\omega}$.}

To quantize the this, we choose
\change{
	\bea
\Phi_{0\omega}= {{\bf a}_{\omega}\over
\sqrt {4\pi|\omega}}e^{-i\omega(t-x)} d\omega+HC\\
\Psi_{0\omega}= {{\bf b}_{\omega}^\dagger\over
\sqrt {4\pi|\omega}}e^{-i\omega(t-x)} b^\dagger +HC
\eea
for $x\rightarrow -infty$.
Since the $\psi_{0\omega} $ is a negative norm
}
incoming mode, it is associated with creation operators.
Also defining operators ${\bf c},~{\bf d}^\dagger$ for the output fields $\Phi_{{\rm
out}\omega}$ in a similar way
\bea
{\bf c}_\omega=\cosh(r(\omega)){\bf a}_\omega+\sinh(r(\omega)) {\bf b}^\dagger_\omega\\
i{\bf d}^\dagger=\cosh(r(\omega)){\bf b}^\dagger_\omega +\sinh(r(\omega)){\bf a}_\omega
\eea
Since $\cosh(r)>1$, we find that the amplitude 
\change{of the
} output $\phi$ field is greater than the
input $\phi$ field\erase{-- This is an amplifier.
a larger output in the $\phi$ outgoing channel.}
\change{. This is an
amplifier since the output in the $\phi$ channel is larger ($\cosh(r)>1$) than
the input in the $\phi$ channel.
}

However, as a quantum system, we note that the outgoing annihilation
operators of $\phi$ field are a linear combination of the ingoing $\phi$
annihilation operator and the ingoing $\psi$ creation operator. This is
a two-mode squeezed state.

\erase{The outgoing annihilation operators obey: 
\bea
{\bf c}_{\omega out}= cosh(r) {\bf a}_{\omega in} +\sinh(r) {\bf
b}^\dagger _{\omega in}\\
{\bf d}_{\omega out} ^\dagger=\cosh(r) {\bf b}^\dagger_{\omega
in}+\sinh(r){\bf a}_{\omega in}
\eea
where I have absorbed the phases into the definitions of $\bf a_\omega$ and $\bf
b_\omega$.} 
\change{We find that
}
\erase{Also $r$ is a non-trivial function of the frequency $\omega$. We
get}
\bea
\cosh(r(\omega))&=&\sqrt{\omega^2+(\epsilon^2+\mu^2)^2\over
\omega^2+(\epsilon^2-\mu^2)^2}\\
\sinh(r(\omega))&=&\sqrt{4\epsilon^2\mu^2\over \omega^2+(\epsilon^2-\mu^2)^2}
\eea

Writing $\ket{0}_{in} $ in terms of the number states for the out 
and the initial state is the vacuum state ${\bf a}_{\omega in}\ket
0={\bf b}_{\omega in}\ket{0}=0$
then the outgoing vacuum state is
${\bf c}_{\omega out}\ket{0}_{out}={\bf d}_{\omega out}\ket{0}=0$ is
\change{ 
\bea
\ket{0}&=&\Pi_\omega (e^{\tanh(r) {\bf c_{\omega}}^\dagger d_\omega^\dagger)}\ket{0}_{out}
\nonumber\\
&=& \sum_{n=0}^\infty\tanh(r)^n \ket {n}_c\ket {n}_d
\eea
}

If we look only at the outgoing $\phi_0$ field, thus tracing out over the
$\psi_0$ field. we get the density matrix for the $\phi$ operators
\bea
\rho_{\phi_{out}}={1\over\cosh(r)^2}\sum_n e^{n ln(\tanh(r)^2}\ket{n}_c\bra{n}_c
\eea
where 
\change{$ln(\tanh(r)^2)=-{\omega\over T_\omega}$.
} \erase{We note that $r$ is a
function of $\omega$.}
For every mode  this is exactly a thermal density matrix with temperature of
\change{$1\over \omega tanh(r(\omega))$
} \erase{This is a thermal
density matrix,}
\change{Each mode thus has an entropy 
}\erase{for
each mode, but the temperature of the modes depends on $\omega$} (i.e., the emission is not
thermal across the modes.)  
The entropy of each mode is then
\bea
S_\omega=- \int \sum_n \tanh(r)^{2n}n
\ln\left(tanh(r)^2\right) 
\eea
Since $|\tanh(r)|<1$, the sum is negative giving a positive entropy
emitted by the amplifier per unit time.

\erase{This entropy does NOT come from some energy states which are mixed by
the non-linear dynamics of the internal degrees of freedom. It is
continuously created by the amplifier out of the incoming vacuum, a zero
entropy, states flowing into the amplifier.}
\change{This entropy does not come
from energy states which are mixed by non-linear dynamics. It is continuously
created by the combination of Bogoliubov entanglement between the $\phi$ and
$\psi$ fields together with ignoring the output of the $\psi$ state from pure
two separate pure incoming states.
} This incoming 
\change{ $\phi$
} mode is {\bf linearly}
separated into positive and negative norm states ( and thus the
description by a non-trivial Bogoliubov transformation). 
It is this model, not that of of a hot lump of coal, which best
describes the entropy emission from a Black Hole as I shall argue in the next
section. 
\change{Of course the entropy of the complete state for all the
degrees of freedom remains zero, in both the lump of coal and the amplifier.
It is the ignorance of the observer of the state of  state of some of the
degrees of freedom that results in the entropy in both the case of the
amplifier and the lump of coal.
} 
As we shall see in the next section, the Black Hole operates very similarly
except that the temperature is not a function of the frequency. Each
\change{exterior observable
}  mode is
thermal, and the temperature of each 
\change{outgoing
} mode is the same. 

\section{Black Hole Entropy Emission}

In the following I am going to look at a Black Hole. To simplify the
analysis, I will make user of a 1+1 dimensional toy model which
we recently published\cite{toy}.

Consider a 1+1 dimensional Schwarzschild Black Hole metric.
\bea \label{toymetric}
ds^2&=& {x-2M\over x}dt^2-{x\over x-2M}dx^2\nonumber \\
&\approx& 16M^2\left[({x\over2M}-1)^2d({t\over 4M})^2-(d{{x\over 2M}-1}))^2\right]; \nonumber \\
&&\hskip 4cm (x~ {\rm near}~0)  \nonumber \\
&\approx& 16M^2\left[d({t\over 4M})^2-(d{x/4M})^2\right] ; ~~{\rm near}~\infty
\eea

This continuous (in $r$)  approximate metric maintains the key
structure of the 2D Schwarzschild metric near the
horizon and at infinity. 
\change{However, instead of the continuous
curvature of the Schwarzschild, it
} has flat Minkowski spacetime in the vicinity of the horizon, and a
different flat spacetime near infinity. It also maintains the Hawking
radiation at infinity.
\bea
ds^2&=&(4M)^2(\rho)^2 d\tau^2-d\rho^2 ;~~0<\rho<{ 1}\\
ds^2&=& (4M)^2 (d\tau^2-d\rho^2);~~\rho>{1}
\eea
where 
\change{$\rho=\sqrt{x/2M-1}$ for $\rho<1$ and is $x$ for
$\rho>1$.
}
Also we have  $\tau=t/4M$. This metric has components which are continuous everywhere
except at $\rho=0$, the horizon. 

This metric looks like the Schwarzschild metric with a
horizon at $\rho=0$, but  near the horizon, ($\rho<1$)  it  is just a form of the  Rindler metric,
a coordinate transformation of a flat Minkowski  spacetime.
Outside $\rho=1$, it is
again just  flat spacetime in the usual Minkowski coordinates, but a different
flat spacetime than  near the horizon. 
\change{It is also 
}a metric which, as we shall see,
has Hawking thermal radiation just as does the Schwarzschild metric.
\erase{While
the Schwarzschild metric has continuous
potentials} 
\change{From $x=2M$ to infinity,
this metric has a
 delta function  curvature scalar at $\rho=1$.
} Since the
metric is conformal flat, massless fields have the same field equations as
they would have in flat spacetime. 
The metric is a glueing together of two flat spacetimes. \erase{, where}
$\rho=1$ is a
curve of constant acceleration 
\change{ as seen from  $\rho<1$
}and is a straight timelike
geodesic for $\rho= 1$ as seen in the metric for $\rho>1.$
Reference\cite{toy} gives  an embedding of this spacetime into a 3-D
globally Minkowski
flat spacetime.

Define 
\change{new null coordinates for this toy model by: 
\bea
U&=&t-x;~~V=t+x\nonumber \\
u&=&\tau-\ln(\rho);~~U=-e^{-u}; ~~U<0\\
\tilde u &=&\tilde\tau-\ln(\tilde\rho);~~U=e^{\tilde u}; ~~U>0\\
\tilde v&=&\tilde \tau+\ln(\tilde\rho);~~ V=-e^{-\tilde v}; ~~V<0\\
v&=&\tau+ln(\rho);~~V=e^{v}; ~~V>0
\eea
The $\tilde{}$ coordinates are behind the horizon and the ``outside"
region (the right Rindler wedge)
is the region $U<0,~~V>0$.}

All of 
\change{
\bea
{dv\over dV},~~{d\tilde v\over d V},~~ {du\over dU}, {d\tilde u\over
d U}
\eea
 are positive everywhere where the $u,~v,~\tilde u,~\tilde v$ are
defined, meaning that time runs the the same direction for
all 
sets of coordinates.
}

The metric 
\change{for
}\erase{in} $\rho<1$ and $\rho>1$ can be written as 
\bea
ds^2 &=&16M^2(-dUdV)\\
&=& -16M^2 e^{v-u}dudv; ~~\rho<1 \\
		 &=&-16M^2 e^{v-u}d\tilde v d\tilde  u{\rm~~\rho>1}
\eea
with equations of motion
\bea
\partial_U\partial_V\Phi=\partial_u\partial_v\Phi=\partial_{\tilde
u}\partial_{\tilde v}\Phi=0
\eea
and with mode solutions
\bea
\phi_L(U)={e^{-i\Omega U}\over\sqrt{2\pi|\Omega|}};~~~\phi_R(V)={e^{-i\Omega
V}\over\sqrt{2\pi\Omega}}\\
\phi(u)=\alpha {e^{-i\omega u}\over \sqrt{2\pi\omega }}\Theta(-U) +\beta
{e^{-i\omega' \tilde u}\over\sqrt{2\pi|\omega'|}}\Theta(-U)
\eea
and similarly for $v$ and $\tilde v$.

The usual Minkowski quantization chooses $\Omega>0$ for the positive
norm modes
proportional to 
\change{$e^{-i\Omega U},~e^{-i\Omega' V}$, to be associated
with the annihilation operators of the Minkowski vacuum. As discussed in
section {\bf C.}, these modes are
analytic in the upper half complex $U$ and $V$ planes, and thus the
annihilation operators are associated with functions which are analytic in the
upper half U and V planes. As shown in {\bf C}, the modes 
\bea
(phi(u,\tilde u)=\alpha e^{-i\omega u}\Theta(-U)+\beta e^{-i\tilde
\omega\tilde u}\Theta(U)
\eea
are analytic in the upper half complex $U,~V$ plane for suitable choice
of $\alpha$ and $\beta$ and relation of $\omega,~\tilde\omega$. In
particular, the modes
  are
}analytic in the  upper half $U$ plane only if
$\tilde \omega=-\omega$ and ${\beta\over \alpha}= e^{-2\pi\omega}$ for
all $\omega$, positive or negative.
In the quantization using the $\phi(u)$ modes, this means that the
positive norm $U$
modes must be two-mode squeezed states when expressed in Rindler
coodinates, just as the $\phi$ modes are  for the amplifier. 
Since, by experiment and prejudice, the Minkowski vacuum has zero energy, this means that
the this two-mode squeeze state near the horizon will have zero energy, and
zero energy flux for $0<\rho<1$.
However for $\rho>1$, the outgoing mode of frequency $\omega$ would be in the
vacuum state, only if one assumed that the $\omega$ modes for the $u$
dependent would have the annihilation operators associated with the
$\omega>0$ modes. After those outgoing ($U$ dependent)  modes flows through $\rho=1$, the
energy-momentum tensor will be non-zero, and one will have a flux of energy
out from the Black Hole which seems to originate at the curvature
delta function, just as was argued in \cite{toy}. 

Why would the delta function curvature change the vacuum state inside
$\rho=1$  to a thermal flux for $\rho>1$?

As we know\cite{notes}, the state of the field as seen by an observer depends on the motion of
the observer. The energy as seen by an observer\cite{wald} is determined by
geodesic point splitting, in which it is the geodesics which determine the
regularization of the field. Outside the horizon, it is the $u$ and $v$ vacua which
have no particles with respect to the geodesics. On the inside, it is the U
and V vacua which have no particles. But at the curvature delta function, the
modes change from being the ones associated with the vacuum state for U and V
to u and v, from a vaccum state to a thermal state.

This model behaves just as does the  amplifier model, with the curvature playing the
role of coupling between the positive and negative modes of the field in the
amplifier.  Here the entaglement is between the outgoing ($U<0$ depedent)
modes 
to the right of the horizon, and the outgoing modes inside the horizon
($U<0$), where we need $\tilde omega=-\omega$.

What of the entropy? Taking the entropy to be the heat flow divided by the
temperature, and the heat flow in both the amplifier and Black Hole being just
the energy, this means that the entropy flux will just be the regularized
energy flow divided by the temperature. Thus the net flow of the entropy out
of the Black Hole will just be the integral of the energy flow divided by the
temperature. The total entropy is not the entropy that resides in the black
hole (with all of the problems  associated with where that entropy is supposed
to hide). Rather it is the total entropy created because of the
two-mode squeezed-state behaviour of the field in the presence of the horizon
\change{and the loss of the observability of the parts of the field behind the
horizons.
}This has a number of implications. The search for where the entropy is stored
in the Black Hole is not a useful endeavour.  The Page curve\cite{page}, for example, is based on the
wrong analogy. However,  the explanation for how the  Geroch heat engine
works-- as being due to
the acceleration temperature by an accelerated observer-- \cite{unruhwald2} still works in
the same way as it does in the Schwarzschild metric. The accelerated box sees the Minkowski vacuum as a thermal state
with the required buoyancy force. 

Throughout the years as I have given lectures on analog Black Holes, one
question almost always arises-- can these analogues say anything about the
"information paradox', and in particular about the entropy of Black Holes? My
answer was always "No", because the temperature in the fluid analog had nothing
to do with the energy. While true, the analogue holes would still produce entropy
by the same mechanism as Black Holes do-- splitting the outputs of two-mode
squeezed states, with one going out to infinity and the other flowing into the
horizon of the analog horizon.  The total entropy emitted will not be related
to the energy emitted, because the temperature has nothing to do with
the energy that has flowed out of the dumb hole, but it will still produce
entropy by the same mechanism as black-holes do. By measuring the entropy or the
two-mode squeezed nature of the created particles by the dumb hole, one can
test the
same mechanism as occurs in a Black Hole. 

Also in the case of 4-D Black Holes, ( or in 2D for massive fields) the modes
that come out of the horizon do not simply flow completely out to infinity as
do the massless modes in 2-D. Instead the angular momentum barrier and the
gravitational barrier cause the low frequency modes ( as seen some distance
away from the horizon) instead get reflected back to the Black Hole. But the
reflected modes carry energy back to the Black Hole. In addition the incoming
vacuum state also gets reflected by the same barriers. This decreases the
negative energy which they carry toward the Black Hole. Thus the mass of the
Black Hole will shrink more slowly, because the mass decreases more slowly.

In conclusion, it seems that  Black Holes themselves have no entropy. They simply continuously create
entropy 
\change{by the loss of observability of the fields behind the horizon
and the 2-mode entanglement between inside and outside parts of the
modes.
}It is like a diner, which does not have a supply of pre-cooked eggs which
they feed to the customers until they are all gone. It is rather like
having a short order cook, who creates
the cooked eggs on demand. 

The analogies to amplifiers are that amplifiers, just like Black Holes emit a
thermal state in each component of the field, if the input was the vacuum state. That thermal
state comes about because of structure of a two-mode squeezed state, with one
of the states (the $\phi$ modes, and the outgoing modes in the Black Hole case).
They differ in that the two parts of the mode, the entaglement between
the  $\phi$ and $\psi$
fields, could in principle be measured. In the case of the black hole,
such a measurement is impossible because because the analogue of the
negative energy $\psi$ field is hidden behind the horizon. 
They also differ in that the amplifier model has a temperature which depends
strongly on the frequency of the mode, while the the black-hole, the
temperature is constant for all the outgoing modes, xince $T=1\over 8\pi M$.

\erase{The input modes into the amplifier model for both fields have the same frequency as the output,
while for the amplifier are the input into the black just at its moment of
formation, with a frequency which is exponentially higher (as a function of
output time)  than the output modes. I.e., the Black Hole transforms input
frequency at a specific time,  into output time delay at specific frequencies,
while the amplifier output is delayed by approximately the same time delay for
all frequencies. Thus for the Black Hole, if one tries to feed back the output
back into the Black Hole, the energy (and entropy)  enters the future horizon and the
singularity with  an approximately fixed delay time (in Kruscal coordinates).
The amplifier amplifies both the energy and the number of photons, by the same
factor, while in the Black Hole case, the energy is exponentially  de-amplified
because of the gravitational redshift, while the number of particles is
amplified. Ie, as always, the analogy is only partial, but I choose it because
in both cases the thermal output is because of  two-mode squeezed
state.}

It is important  that in both cases, \change{(amplfier and Black Hole),}
just  as in  Hawking's
original derivation was for a linear field.  There is no "mixing" of modes which would
require  non-linear processes to produce the  entropy, as happens within a
lump of coal. 

The above depends critically on the idea that General Relativity is 
good approximation to the true theory of gravity near the horizon. 
If, for example, string theory is a better approximation to the reality of
gravity, then the string theory arguments about the relation between the
entropy of Black Holes and, for example, D-Branes explanations,  may be closer to reality. In
that model,
perhaps Black Holes do have entropy. \erase{There one must go through some more or less
convoluted arguments to obtain the temperature of a Black Hole.}

\section{ Acknowledgements}
I thank the IQSE and Hagler Institute at the Texas A\&M University for
support while goading me to think more deeply about the Black Hole
evaporation and the relationship to quantum optics. I especially thank the
M Aspelmeyer who invited me to spend some time at the IQOQI and the University
of Vienna where this paper was mostly written.
I thank the UBC administration who came through with support when NSERC
cancelled my funding.
Finally, I  thank my wife, Patricia Unruh,  whose support and help has been  crucial during my trips
and during my writing of this paper.

\section{END NOTES}
\subsection{Entropy}
The density matrix is
\bea
\rho\propto\int \sum_{n_\omega}e^{n_\omega\omega}
\ket{n_\omega,\tilde\omega}\bra{n_\omega,8\pi M \tilde\omega}d\tilde\omega
\eea
where $\omega=8\pi M \tilde\omega$ and $\tilde\omega$ is the energy of mode.
The sum over the probabilities $\sum_{n_\omega} e^{n_\omega\omega} ={1\over
{1-e^{-\omega}}}$ is precisely the square of the factor in front of the mode
\bea
J_(t,x)_{\tilde\omega}&=& {i\over
4\pi|\tilde\omega|}\left(\phi(t-x)^*_{\tilde\omega}\partial_x\phi(t-x)\phi(t-x)_{\tilde\omega}\nonumber\right.
\nonumber\\
&~&~~~~~\left. -\phi(t-x)_{\tilde\omega}\partial_x\phi(t-x)\phi(t-x)_{\tilde\omega}\right)
\eea
the flux of the Klein Gordon norm for the mode.
Thus the flux of energy is just 
\bea
<{\rm J}_E>=\int\tilde \omega {\rm J_{\tilde\omega}}(t,x) d\tilde\omega
\eea
and the Entropy flux is
\bea
<{\rm J}_S(t,x)>= \int 8\pi M \tilde\omega {\rm J_{\tilde\omega}}(t,x)
d\tilde\omega =8\pi M <{\rm J}_E>
\eea 
which, with $T= {1\over 8\pi M}$, the Hawking temperature, is just
\bea
<{\rm J}_S(t,x)>= <{\rm J}_E>/T
\eea
\change{
\subsection{Two-Mode Squeezed State}
}

\change{
Consider a linear system with two degrees of freedom. It is assumed to have
an input and an output each with two degrees if freedom. input annihilation
operators are $\bf a$ and $\bf b$, while the output degrees of freedom
are represented by the annihilation operators
$\bf c$ and $\bf d$. We will work in the Heisenberg reresentation in which
the operators are dynamic and the states are static.
}

\change{
Let us assume that the ouput operators, which are linearly related to the
input operators, are given by
\bea
c=\alpha {\bf a}+\beta {\bf b}^\dagger;~~~ d=\gamma {\bf b}+\delta {\bf
a}^\dagger
\eea
a relation which is also called a Bogoliubov transformation.
Then the commutation relations are
\bea
\left[{\bf a,a}^\dagger\right]&=&\left[{\bf b,b}^\dagger\right]=\left[{\bf c,c}^\dagger\right]=\left[{\bf
d,d}^\dagger\right]=1 \nonumber \\
\left[{\bf a,b}^\dagger\right]&=&\left[{\bf
a,c}^\dagger\right]=\left[{\bf a,d}^\dagger\right]
\nonumber\\
&=&\left[{\bf b,c}^\dagger\right]=\left[{\bf b,d}^\dagger\right]=\left[{\bf c,d}^\dagger\right]=0\\
\eea
and the Hermitian conjugates of these equation. the commutators
between any two annihilation operators are all equal to 0.
}

\change{  
By chooosing the phase of the operators appropriately, all of $\alpha,
beta,\gamma,\delta$ can be made real.These commutation relations demand that
\bea
\alpha^2-\beta^2=\gamma^2-\delta^2=1
\eea
which implies that $\alpha^2,~\gamma^2$ are both greater than or equal to 1. 
Also we have
${\beta^2\over\alpha^2}={\delta^2\over\gamma^2}$, or 
\bea
\alpha^2=\gamma^2; ~~\beta^2=\delta^2
\eea
The initial vacuum state is
\bea
{\bf a}\ket{0,0}_{in}={\bf b}\ket{0,0>}_{in}=0
\eea
while the outgoing vacuum state is
\bea
{\bf c}\ket{0,0}_{out}={\bf d}\ket{0,0}_{out}=0
\eea
In terms of the ``in" operators and vacuum, we have,
in the Heisenberg representation, the state of the system does not
change, so,
if one starts with the ``in" vacuum, the final state will be the same. However,
since the ``out" operators are not the same as the ``in", we use the
``out" operators
to define the ``out" quanta. Then $\ket{0,0}_{in}=
F(c^\dagger,d^\dagger)\ket{0,0}_{out}$
with
\bea
F(c^\dagger,d^\dagger)={\mathfrak N}e^{-{\beta\over\alpha}{\bf c}^\dagger{\bf d}^\dagger}
\eea
where $\mathfrak N$ is a normalization constant.Defing ${\mathfrak
T}={\beta\over \alpha}$ we have
\bea
\ket{0,0}_{in}={\mathfrak N}\sum_{n=0}^\infty{ \mathfrak T}^n \ket{n,n}_{out}
\eea
}

\change{
Now, if we are looking only at the $\bf c$ degree of freedom, we get the
density matrix
\bea
\rho_c&=& {\mathfrak N}^2\sum_n {\mathfrak T}^{2n}(\ket{n}\bra{n})_c
\nonumber\\
&=&{\mathfrak N}^2\sum_n e^{n\ln({\mathfrak T}^2)}(\ket{n}\bra{n})_c
\eea
which is just the themal density matrix with $-\ln({\mathfrak T}^2)={E\over
T}$ with $T$ the temperature, and $E$ the energy difference between the quanta
energy levels of the $c$ degree of freedom. (Note that $\ln({\mathfrak T}^2)$
is less than 1 so the logarithm is less than 0.)
This is just the density matrix of  a thermal state. 
}

\subsection{Minkowski Rindler modes}

\change{Consider the null coordinates, $U,~u,~\tilde u$, as  defined in
the main body,  and define
the complex $U=U_R+iU_I$. A positive norm
Minkowski normalised mode is
\bea
\phi_\Omega\propto {e^{-i\Omega U}\over\sqrt{2\pi|\Omega|}}; ~~~\Omega>0
\eea
These are a complete set of positive norm modes, and, together with their
complex conjugates, they form a complete set of massless "right travelling"
modes.
}

\change{All of these postive norm modes are such that they are analytic
in the region with positive  imaginary values $U_I$ as they go to 0 as $U_I\rightarrow
\infty$. 
}

\change{Now consider the complex function $-U^{-i\omega}$. This function
has a branch cut starting at U=0. If we take this branch cut going
from $U=0$ to $U=-i\infty$, this function is analytic (no singularities,
no poles) for positive imaginary part.
Thus, this function will be writeable in terms of the postive norm functions
$\phi_\omega$ for $\omega>0$ for all values of $\omega$ positive or
negative.  }

\change{We can write this in the regime $U_I>=0,~U<0$
\bea
U^{i\omega}\propto e^{-i\omega\ln(-U)} =e^{-i\omega u}
\eea
Since it analytic for $U_I>0$, as we go along the real $U$ axis we must
go around $U=0$ in the upper $U$ plane, where $Re i\omega\ln(-U)$ for
$U<0$  goes to $e^{-{\pi\omega}\ln(U}$ for $U>0$. Thus the postive norm
mode is proportional to 
\bea
\psi_\omega= e^{-i\omega\ln(-U)\Theta(-U) +e^{-\pi\omega}
e^i\omega\ln(U)} \Theta(U)
\\
=e^{-i\omega u}\Theta(-U) +e^{-\pi\omega}e^{i\omega(-\tilde u)}
\eea
But $e^{-i\omega u}$ is a complete set of positive norm modes ($u$ is a
coodinate going from $-\infty$ at $U=-0$ to $U=-\infty$, and its complex
comjugate is a complete set of negative norm modes in the same range.
Similarly $e^{-i\omega\tilde u}$ is a complete set of negative norm
modes for $\tilde u=-\infty$ at $U=0$ to $\tilde u=\infty$ at
$U=\infty$. These are the Rindler null coordinates, defined so that as
bot $u$ and $\tilde u$ increase, they correspond to $\tau$ increasing.
Ie, a positive norm Minkowski mode is a combination of the positive
Rindler $u$ mode  outside the horizon $V=0$ and a negative norm Rindler
mode inside the horizon ($V<0$). Ie, the set of positive norm Minkowski
modes are a conerent ($\omega$ is the same for both sides) sum of
positive norm Rindler modes outside the horizon if $\omega>0$ and
negative norm Rindler mode if $U>0$ inside the horizon. The positive
norm mode Rindler  has an exponentially  greater amplitide, than the
negative norm on the opposite side of the horizon (and vice-versa if
$\omega<0$). Ie, positive norm Minkowski modes are two-mode squeezed
states as far as the Rindler modes are concerned, one of the modes
non-zero outside the horizon and the other inside the horizon, where it
is unmeasureable by an observer outside the horizon.
See \cite{notes} for a more detailed exposition. Beware as  I did not
then 
know what a two-mode squeeezed state was.
}

\begin{thebibliography}{99}
\change{\bibitem{Hawking} S.W.~Hawking "Black Hole
explosions?". Nature. {\bf 248} 30-31
(1974)
}
\bibitem{WeighingofEnergy} G. Opat, W.~G. Unruh "The Bohr-Einstein
Weighing of Energy debate" Am. J. Phys. {\bf 47}, 743 (1979)
\bibitem{unruhwald} W.~G.~Unruh and R.~M.~Wald, "What happens when an accelerating observer detects a
Rindler particle", Phys.~Rev.~D {\bf 29}  1047-1056 (1984)
\bibitem{notes} W.~G.~Unruh.``Notes on Black Hole evaporation",
Phys.~Rev.~D{\bf 72} 870-892 (1974)
\bibitem{bhamp} W.~G.~Unruh "Quantum Noise in Amplifiers and
Hawking/Dumb-Hole Radiation as Amplifier Noise", arXiv:1107.2669v2
published in ed: L.C.B.~Crispino, V.M.S.~Cardoso, S.~Liberati,
E.S.~Oliveira, M.~Visser "Analogue Spacetimes--the
first thirty years: Proceedings of the Amazonian workshop II
on Analogues"  VLF Editorial, Editora Livraria da Fisica, Sã
Paulo (2013) (Out of print and not accessible in libraries.)
\bibitem{schwablthirring} F.~Schwabl,
W.~Thirring on "Quantum Theory of Laser Radiation" in {\bf
Ergeb.~d.~Exakten Naturwissenschaften},   {\bf 36},
219,  Ed. G.~Hoeler  (Springer 1964) and M.~Scully, W.~Lamb
Phys. Rev. {\bf 159}, 208-226 (1967)
B.R.~Mallow, R.J.~Glauber "Quantum Theory of Parametric
Amplification.II" Phys Rev {\bf 160} 1097-1108 (1967)
\bibitem{wald}	S.~Hollands, R.M.~Wald ``Quantum Fields in Curved Spacetime"
Physics Reports {\bf 574} 1-35 (2015)
\bibitem{2modetherm}  S.M.~ Barnett and P.L.~Knight, 
``Thermofield analysis of squeezing and statistical mixtures in quantum optics",
J.~Opt.~Soc.~Am. {\bf B 2}, 467 (1985), doi:10.1364/JOSAB.2.000467 or .
B.~Yurke and M.~Potasek, Obtainment of thermal noise from a pure quantum
state, Phys.~Rev.~{\bf A 36}, 3464 (1987), doi:10.1103/PhysRevA.36.3464.	
\bibitem{tunnel} Maulik K. Parikh, Frank Wilczek ``Hawking Radiation as
Tunnelling" Phys.Rev.Lett {\bf 85} 5042-5045 (2000)
\bibitem{string}Sumit R. Das ``Black Hole Entropy and String Theory"
arXiv:hep-th/9602172 or M. Duff, Hong Lu, C. Pope ``The black branes of
M-theory" Physics Letters {\bf B 382} 1-2, 73-80.(1996)
\bibitem{toy} R.~Schuetzhold, W.~Unruh ``Toy model of Black Hole evaporation"
Comptes Rendus Physique {\bf 25} 1-10 (2025)
\bibitem{page} D.~Page  "Information in Black Hole Radiation". 
Phys.~Rev.~Lett.~{\bf 71} 3743-3746 (1993) ; arXiv:hep-th/9306083
\bibitem{unruhwald2}W.G.~Unruh., R.M.~Wald ``Entropy bounds,
acceleration radiation, and the generalized second law" 
Phys.~Rev.~{\bf D27} 2271-2276 (1983).
\end{thebibliography}
\end{document}